\documentclass[journal]{IEEEtran}
\usepackage{cite}
\usepackage{amsmath,amssymb,amsfonts}
\usepackage{algorithmic}
\usepackage{graphicx}
\usepackage{textcomp}
\usepackage[percent]{overpic}
\usepackage{amsmath,amsthm,amssymb}
\usepackage{amssymb, amsthm}
\usepackage{amsfonts}
\usepackage{amsmath}
\usepackage{graphicx}
\usepackage{xcolor}
\DeclareGraphicsExtensions{.pdf,.png,.jpg}
\newtheorem{theorem}{Theorem}[section]

\newtheorem{lemma}[theorem]{Lemma}
\newtheorem{coroll}[theorem]{Corollary}

\newtheorem{deff}[theorem]{Definition}

\def\BibTeX{{\rm B\kern-.05em{\sc i\kern-.025em b}\kern-.08em
    T\kern-.1667em\lower.7ex\hbox{E}\kern-.125emX}}
\begin{document}
\title{Multivaluedness in Networks: Theory}  
\author{\scalebox{1.25}{Micha\"el Anton(ie) van Wyk, 
}
\linebreak\linebreak \scalebox{0.95}
{University of the Witwatersrand, Johannesburg, South Africa,} \linebreak
\scalebox{0.95}{and}
\linebreak
\scalebox{0.95}{City University of Hong Kong, Hong Kong, 
SAR of the PRC}

\thanks{
\vspace*{2mm}

Financial Support Acknowledgment: This work was supported in part 
by the Carl and Emily Fuchs Foundation's Chair in Systems and 
Control Engineering at the University of the Witwatersrand, 
Johannesburg, South Africa.

\vspace*{3mm}
}

}

\maketitle

\begin{abstract}
An unexpected and somewhat surprising observation is that  
two counter-cascaded systems,\footnote{In Fig.~\ref{Fig_1} the 
systems $N\circ T$ and $M$ are \emph{counter-cascaded systems} 
with common input $u$.} given the right conditions, can exhibit 
multivaluedness from one of the outputs to the other.  The main 
result presented here is a necessary 
and sufficient condition for multivaluedness to be exhibited 
by counter-cascaded systems using the novel notions of 
\emph{immanence} and its opposite, \emph{transcendence}, 
introduced here.  Subsequent corollaries provide further 
characterization of multivaluedness under specific conditions.

As an application of our theoretical results, we demonstrate 
how these aid in the structural complexity reduction of complex 
networks.
\end{abstract}

\begin{IEEEkeywords}
Big data, complex networks, functional uniformization, immanence, 
multivaluedness, multivalued mapping, multivalued relation, neural 
networks, network analysis, network science, networked  systems, 
nodal rationalization, structural reduction, transcendence.
\end{IEEEkeywords}


\section{Introduction}
\label{sec:introduction}
The area of Network Analysis and Complex Networks has rapidly 
expanded into a very active and vibrant field of research, with 
ever more fundamental theoretical results and novel applications 
being reported.  
In order to give a glimpse of the diverse nature of the objects 
of study, i.e. complex networks, note that, size-wise, real 
life networks range from a few nodes to billions of nodes 
and beyond.  Structure-wise, they range from highly homogeneously 
structured networks through to amorphously unstructured and 
even randomly structured networks.  Character-wise, they vary 
from uniformly cooperative or competitive to heterogeneously 
mixed cooperative and competitive factions contained within.
Furthermore, the mathematical descriptions of nodes in a 
complex network range from uniform (identical) in some 
networks, to diverse (different) in others.  For these 
reasons, graph-theoretical methods are indispensable for 
description and analysis of network problems.

In the literature, the meaning of the term ``network analysis'' 
is rather diverse.  Of particular interest to us here, is the 
extended definition of Zaidi \cite{Zaidi2010}, namely that it encapsulates 
the study of theory, methods and algorithms applicable to 
graph-based models representing interconnected real-world 
systems.  From this perspective, the collection of interconnected 
elements of a finite element analysis of a distributed structure 
or physical field and a complex interconnection of nonlinear 
dynamical systems are instances of complex network analyses 
\cite{XiangChenRenChen2019}\!\!\!\cite{Wu2010}, the former 
undirected and the latter directed.  Both an excellent account 
of the theory and overview of current research directions in 
complex networks, can be found in \cite{ChenWangLi2015}.  

Even though complex networks might not always have explicit 
inputs (causes) and outputs (effects), there are always 
internal (i.e. local) inputs and outputs of interest 
when considering a single node or a collection of nodes. 
A deeper understanding of the global behavior and dynamics 
of a complex network usually requires a deeper understanding 
of the mechanisms of behavior at a more detailed level in the 
network.  For this reason, oftentimes it requires one to relate 
two effects brought about by the very same global or local 
cause, in order to gain deeper insight.  In this paper we 
study this aspect in detail.

The outline of this paper is as follows:  Section~\ref{sec:Theory}
presents a simple yet powerful theoretical result that gives 
a necessary and sufficient condition under which two different 
(sets of) effects $v\in V$ and $x\in X$, produced by the common 
cause $u\in U$, are to be related by a well-defined mapping.
Due to the minimal underlying assumptions and the simplicity 
of the set-theoretic argument used, the results derived is 
very general.  Next, several consequences of the main result 
are addressed.  By virtue of an example, Section~\ref{sec:Applic} 
demonstrates the use of these results when applied to 
structural reductions in complex networks that we will 
call functional uniformization and nodal rationalization. 
The conclusion follows in Section~\ref{sec:Conclu}.

\section{Theory: Immanence versus Transcendence}
\label{sec:Theory}
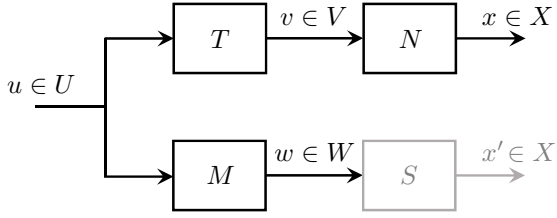
\begin{figure}
\vspace*{-3mm}
\begin{center}
\begin{overpic}[width=0.40\textwidth]{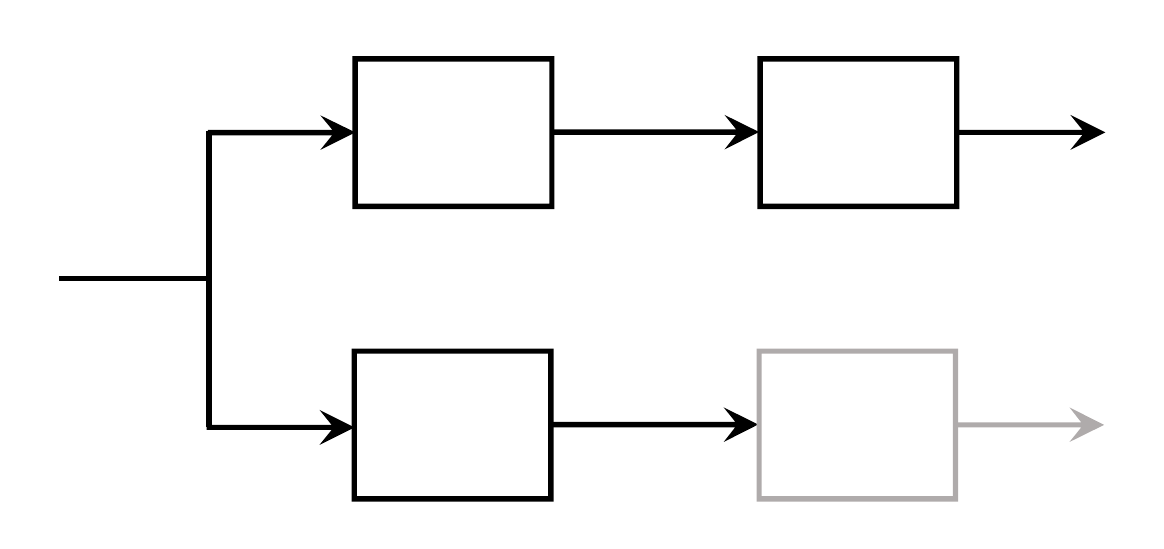}  
 \put (0,25.7)    {$u\in U$}
 \put (50,38.2) {$v\in V$}     \put (87,38.2) {$x\in X$}
 \put (37.3,34) {$T$}          \put (71.8,34) {$N$}
 \put (49,13.2) {$w\vspace*{-.6mm}\in\vspace*{-.6mm} W$}
 \put (87,13.3) {\color{gray}$x'\! \in X$}
 \put (36.5,9)  {$M$}          \put (72,9)    {\color{gray}$S$}
\end{overpic}
\end{center}
\vspace*{-5mm}
\caption{Two counter-cascaded paths with the common cause, $u\in U$.} 
\label{Fig_1}
\end{figure}

In order to provide a definite and concrete 
context\footnote{Concreteness, here, does not restrict 
generality.} for the presentation and discussion that follow, 
we consider complex networks consisting of complex configurations 
of nonlinear (dynamical) systems.  In such networks, we will 
study occurrences of \emph{counter-cascaded} systems, i.e.\ 
configurations of the kind shown in Fig.~\ref{Fig_1}.  
Generally, $U$, $V$, $W$ and $X$ can be very general 
sets, with  $M: U\to W$, $T:U\to V$ and $N:V\to X$ mappings.  
For the selected context, unless stated otherwise, these 
mappings are nonlinear operators with  
domains and ranges being subsets of 
real vector spaces; typically $T$ is a nonlinear operator 
describing some nonlinear system, with $M$ a nonlinear operator 
describing either yet another nonlinear system or an input 
ancillary system, and $N$ is a nonlinear operator representing 
some output ancillary system.  
In some applications $M$ might be the identity operator, as 
does $N$.  For the purpose of our presentation here, $N$ is 
redundant because it can be absorbed into $T$ by replacing $T$ 
with $N\circ T$.  However, for applications of this work 
in other areas, it has a distinct and explicit purpose, as will 
be reported on in the future.  Finally, $S\subset W \times X$ 
usually represents a multivalued function, strictly called a 
relation.

In order to simplify the notation used here, we will not 
distinguish a system from its mathematical representation 
and thus use the same symbol for both.  Furthermore, in order 
to retain the generality of the results presented, we will 
talk of ``mappings'' rather than ``operators'' as required by 
the nonlinear systems context.

\begin{deff}  \emph{\textbf{(Immanence, Transcendence)}}\ 
\label{InfStrucPres} 
In Fig.~\ref{Fig_1}, the mapping $T$ is called \emph{immanent with respect (or relative) to the ordered pair of mappings} $(M, N)$ if, 
for every $w\in M(U)$ there exists an\footnote{If such $x$ exists, 
then it is unique.  To see this assume that at least two such 
elements $x_1$ and $x_2$ exist implying that $N^{-1}(x_1)\bigcap 
N^{-1}(x_2) \ne\emptyset$. Applying $N$ to this intersection 
immediately yields $x_1=x_2$.\vspace*{1.2mm}} $x\in N(T (U))$ 
such that $T(M^{-1}(w)) \subseteq N^{-1}(x)$.

If $T$ is not immanent with respect to $(M,N)$, then it is called 
\emph{transcendent} with respect to $(M,N)$.
\end{deff}
\bigbreak

%


\noindent\emph{Notes.} 
\begin{enumerate}
%
\item[a.] For the sake of conciseness, we will sometimes use 
  the statement ``$T$ is $(M,N)$-immanent'' as an abbreviation 
  of the statement ``$T$ is immanent with respect to 
  $(M,N)$'' and similarly for statements about transcendence.
\item[b.] For a general complex network, in order for two nodes 
  to be analyzed for immanence or transcendence, their inputs 
  have to be connected together.
\item[c.] Since collections of nodes can be clustered to form 
  supernodes, which are themselves nodes, this definition and 
  all subsequent results apply to supernodes without explicit 
  further mention.
\end{enumerate}\vspace*{1mm}

Assuming $M$ and $N$ generally to be many-to-one, we are now 
in a position to state and prove the main result:

\bigbreak
\begin{theorem}  \emph{\textbf{(Well-Defined Mapping)}}\ \label{ModOpTheor} 
The mapping $T$ is immanent relative 
to $(M,N)$ if and only if $N\circ T\circ M^{-1}$ is well-defined 
(i.e.\ single-valued).\footnote{Here, $M^{-1}$ denotes the preimage 
of $M$ and $I$ denotes the identity mapping.}
\end{theorem}

\noindent\emph{Note.}\ Before proceeding with the proof, 
first observe that for each element $u\in U$, there exist elements 
$w_u := M(u)$ and $x_u := N(T(u))$.  
Next, we associate $w_u$ and $x_u$ by writing $x_u = S(w_u)$ for 
every $u\in U$.  This can be compactly expressed as 
$S:=N\circ T\circ M^{-1}$.  Here, $S$ defines a relation. 
If for every pair of distinct 
elements $u_1, u_2 \in U$ we have that $w_{u_1} = w_{u_2}$ implies 
that $x_{u_1} = x_{u_2}$, then $S$ is well-defined. 

\begin{proof}
We first prove the ``only if'' part.  Suppose that $T$ is 
immanent with respect to $(M, N)$.  Now, 
if $S$ is not well-defined, then there exist at least two distinct 
elements $u, u'\in U$ such that $M(u) = M(u')$ but $x\!:=\! N(T(u)) 
\ne N(T(u')) \!=:\! x'$. This contradicts the consequence of 
immanence, namely that $N(T(u)) = x$ 
for all $u\in M^{-1}(w)$ and consequently $S$ is well-defined.

Conversely, to prove the ``if'' part, suppose $T$ is transcendent 
with respect to $(M,N)$. Then, for some 
$w\in W$, there are distinct elements $u, u' \in M^{-1}(w)$ for which 
$x := N(T(u)) \ne N(T(u')) =: x'$, implying  that $S$ is not 
well-defined because $S(w) = x$ and $S(w) = x'$ and yet $x \ne x'$.  
This concludes the converse via the contrapositive and completes 
the proof.
\end{proof}
\bigbreak

An equivalent statement of this result follows:

\bigbreak
\begin{theorem}  \emph{\textbf{(Multivalued Relation)}}\ \label{MulValRel} 
The mapping $T$ is transcendent  
relative to $(M,N)$ if and only if $N\circ T\circ M^{-1}$ is 
not well-defined (i.e.\ multivalued).
\end{theorem}
\bigbreak

To our knowledge this result, identifying all those situations 
when the outputs of two counter-cascaded subsystems are 
functionally related (as well as when not), is a novel 
result.\vspace*{1.0mm}

Some immediate consequences of Theorem~\ref{ModOpTheor} now follow.
\bigbreak
\begin{coroll} \emph{\textbf{(Existence of a Unique Faithful Model)}}\ 
\label{ModelingError} 
For a given mapping $T$, a unique faithful \emph{model} or 
\emph{modeling mapping} $S$ exists if and only if $T$ is 
immanent with respect to $(M,N)$.
\end{coroll}
\bigbreak
If $T$ is immanent with respect to $(M,N)$, then there 
exists a unique mapping $w \mapsto S(w)$ which yields a 
unique faithful model of $T$, as perceived through $M$ 
and $N$, that is,  
$S(w) = N\circ T\circ M^{-1}(w)$ for every $w\in W$.  Even 
though these modeling problems are exactly solvable, in 
principle, it might happen that the prescribed class of 
models does not contain $S$, in which case the best that 
can be achieved is to choose the ``best'' approximating 
model $S_\mathrm{opt}$ from the class prescribed, based on 
some optimization criterion.

On the other hand, if $T$ is transcendent with respect 
to $(M,N)$, then there exists a (multivalued) relation 
$S \equiv N\circ T\circ M^{-1}$, which \emph{cannot} be 
described by any mapping, whatsoever, and hence \emph{no} 
faithful model exists.  Therefore, the only alternative 
remaining is to find the ``best'' approximating mapping 
$S_\mathrm{opt}$ to the relation $S$, based on some criterion.
\bigbreak
\begin{coroll} \label{ComprehensiveCor} 
Let $M$, $T$, $N$ and $S$ be as depicted in 
Fig.~\ref{Fig_1}.\vspace*{2mm}
\begin{enumerate}
\vspace*{-1.5mm}
\item[a.] If $M$, $N$ are given and $T$ assumes the 
  \emph{canonical form} $T = F\circ M$ for some fixed 
  $F: W\to V$, then $T$ is $(M,N)$-immanent.
\item[b.] If $M$ is bijective then every $T$ is $(M,N)$-immanent 
  for every $N$.
\item[c.] If $M$ is many-to-one, $T$ is not of canonical form 
  and $N$ is one-to-one, then $T$ is $(M,N)$-transcendent.
\item[d.] If $M$ is one-to-one, $N$ is many-to-one and $T$ is 
  $(M,N)$-immanent, then the mapping $S$ is many-to-one.
\item[e.] Given a fixed mapping $S$, if there exists a $T$ 
  satisfying $N\circ T= S\circ M$, then $T$ is $(M,N)$-immanent 
  for every mapping $N$.
\end{enumerate}
\end{coroll}
\bigbreak
\noindent\emph{Notes.} 
\begin{enumerate}
\item[a.] If $M$ is many-to-one and $T$ is \emph{not} 
of the canonical form $T = F\circ M$, then the $(M,N)$-immanence 
of $T$ depends on the choice of $N$.
\item[b.] In Corollary~\ref{ComprehensiveCor}(e), if $N$ is 
invertible, then the expression $T = N^{-1}\circ S\circ M$ 
gives an explicit formula for $T$.  However, if $N$ is not 
invertible then $T$ 
satisfies the expression $N^{-1}\circ N\circ T = N^{-1}\circ S
\circ M$ which is generally not solvable for $T$ since $N^{-1}
\circ N \ne I$.  So, unless additional information about $T$ 
is available, we can merely test candidate mappings $T$ to 
determine if they satisfy this expression. .... In the latter 
case, is $T$ $(M,N)$-immanent or $(M,N)$-transcendent?
\end{enumerate}
\vspace*{2mm}

Now, a little thought reveals the following to be true for 
configurations similar to that shown in Fig.~\ref{Fig_1}, but 
with additional exogenous inputs entering:
\bigbreak
\begin{lemma}  \emph{\textbf{(Resolution of Exogenous Inputs)}}\ 
\label{Incorp} 
Suppose that along some of the paths considered to yield the 
mappings 
$M$ and $T$, there are additional causes entering. Then this 
configuration can be transformed to that in Fig.~\ref{Fig_1} 
by augmenting the input space $U$ with a direct sum component for 
each additional exogenous input entering. After this transformation, 
proceed as before to define the mappings $M$ and $T$.
\end{lemma}
\bigbreak

\noindent\emph{Note.} Since, by assumption 
$N$ does \emph{not} take the input $u\in U$ directly, this 
lemma does \emph{not} apply to it.  In more general situations 
where $N$ shares an input with $M$, replacing $T$ with $I\oplus T$, 
the structure depicted in Fig.~\ref{Fig_1} applies once more.  The 
symbol $\oplus$ represents the direct sum binary operation. 
\vspace*{1mm}

In the case of counter-cascaded systems, there are two possible 
directions to be considered for immanence or transcendence.  
The next definition expands on the previous definition to 
cover both possibilities.  For this, $N$ will effectively be 
removed by choosing it to be the identity mapping.

\bigbreak
\begin{deff}  \emph{\textbf{(Bi-immanence, Bi-transcendence)}}\ 
\label{BiInfStrucPres} 
In Fig.~\ref{Fig_1}, 
if $T$ is $(M,I)$-immanent and $M$ is $(T,I)$-immanent, then $T$ and 
$M$ are called \emph{bilaterally immanent} or \emph{bi-immanent}. 
Similarly, if $T$ and $M$ are $(M,I)$- and $(T,I)$-transcendent, 
respectively, then $T$ and $M$ are called \emph{bilaterally 
transcendent} or \emph{bi-transcendent}.
\end{deff}
\bigbreak

\noindent\emph{Note.} Considering the two possible directions 
along two counter-cascaded systems, in principle, all of the 
following four cases are possible: immanent-immanent (I-I), 
immanent-transcendent (I-T), transcendent-immanent (T-I) 
and transcendent-transcendent (T-T).  For the case I-I 
the mapping relating the outputs is a bijection while, 
for the case T-T, there is no mapping that relates the 
two outputs in either direction. The case I-T implies 
that such a mapping exists in one direction but not in 
the other; similarly for the case T-I.

\vspace*{2mm}
\noindent\emph{Discussion.} An important insight obtained 
from the above theoretical development is that, for cases 
with the mappings $T$ and $M$ \emph{given}, but with $T$  
transcendent with respect to $(M,I)$, the only  way to resolve 
this situation, if at all possible, is to design an 
appropriate output ancillary mapping $N$.  If this proves 
to be impossible, then the above theorems imply that more 
design freedom is required.  For example, we can allow $M$ 
to be engineered or redesigned in an attempt to find a pair 
$(M,N)$ rendering $T$ to be immanent. If still not successful, 
$M$ could be fixed and $T$ be redesigned.  If still not 
successful, then no choice remains other than a complete 
redesign.  However, at any stage we could settle for 
approximation, then knowing that a solution does not exist 
which implies nonzero approximation error as a consequence 
of the prevailing transcendence.

In some applications, however, it happens that the mapping $S$ is 
given instead, and the mapping $T$ then follows as a consequence 
of $S$ and the particular application's context and constraints. 
For such cases, 
it might be necessary to adapt either the context and/or 
the constraints, in order to obtain a $T$ that is 
$(M,N)$-immanent with $M$ and $N$ implied by the 
application's context and constraints.

\section{Application: Nodal Rationalization}
\label{sec:Applic}
We will now apply the results of the previous section 
to structural reduction in complex networks.
We start by introducing the necessary terminology.  The 
process of expressing node mappings in factored form, with the 
right-most factors chosen from as few as possible unique 
ones, will be referred to as \emph{functional uniformization}. 
Furthermore, the process of minimizing the number of nodes in 
a functionally uniformized network, by merging as many nodes 
as is possible to share common right-most factors and (node) 
inputs, will be termed \emph{nodal rationalization}.  
This form of structural transformation of a network results 
in a reduction in the number of nodes, with each of the 
newly formed nodes having either multiple inputs or multiple 
outputs or both.

\begin{figure}[h]
\vspace*{-3mm}
\begin{center}
\begin{overpic}[width=0.27\textwidth]{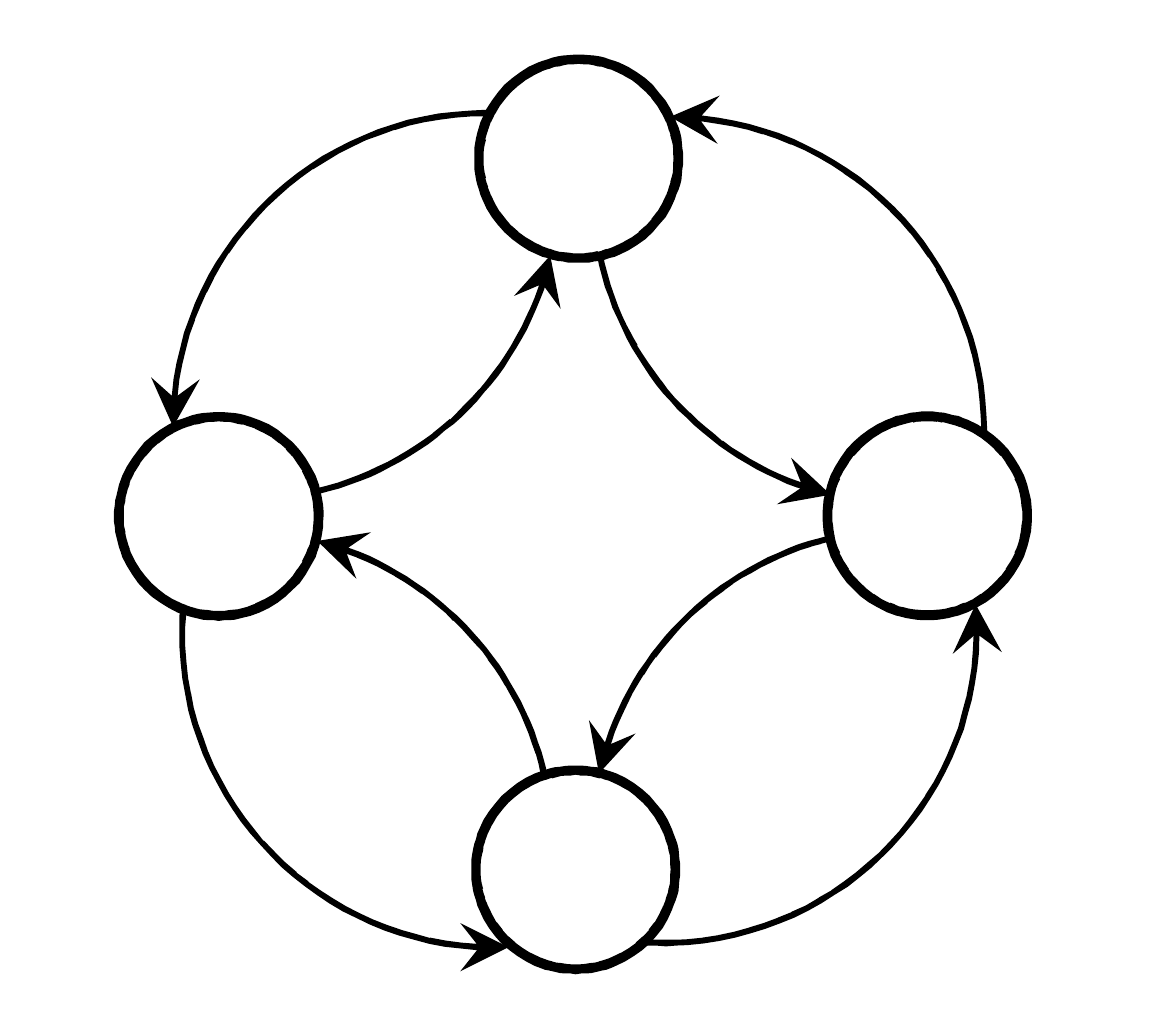}  
 \put (46.9,72.1)    {$A$}
  \put (77.4,40.6)    {$B$}
  \put (47.1,9.8)    {$C$}
 \put (15.8,40.7)    {$D$}
\end{overpic}
\end{center}
\vspace*{-5mm}
\caption{A four-node complex network immediate neighbor 
interaction.} 
\label{Fig_2}
\end{figure}
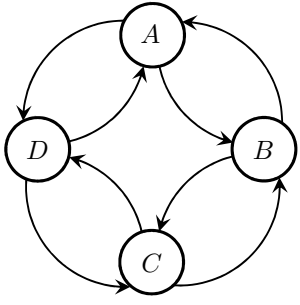

Consider the simple yet general four-node network with 
immediate neighbor interaction, shown in 
Fig.~\ref{Fig_2}. We will use the same symbol to present 
both the node and the mathematical mapping describing its 
behavior, i.e. its mathematical description.  For example, 
$A$ identifies the upper-most node in Fig.~\ref{Fig_2}; 
it also represents the mathematical mapping $A(\cdot,\cdot)$
that describes this node's behavior.

For clarity of presentation, we will relate back to earlier 
theoretical results by using the nonlinear systems representation 
employed in the previous section.

\begin{figure}
\vspace*{-3mm}
\begin{center}
\begin{overpic}[width=0.33\textwidth]{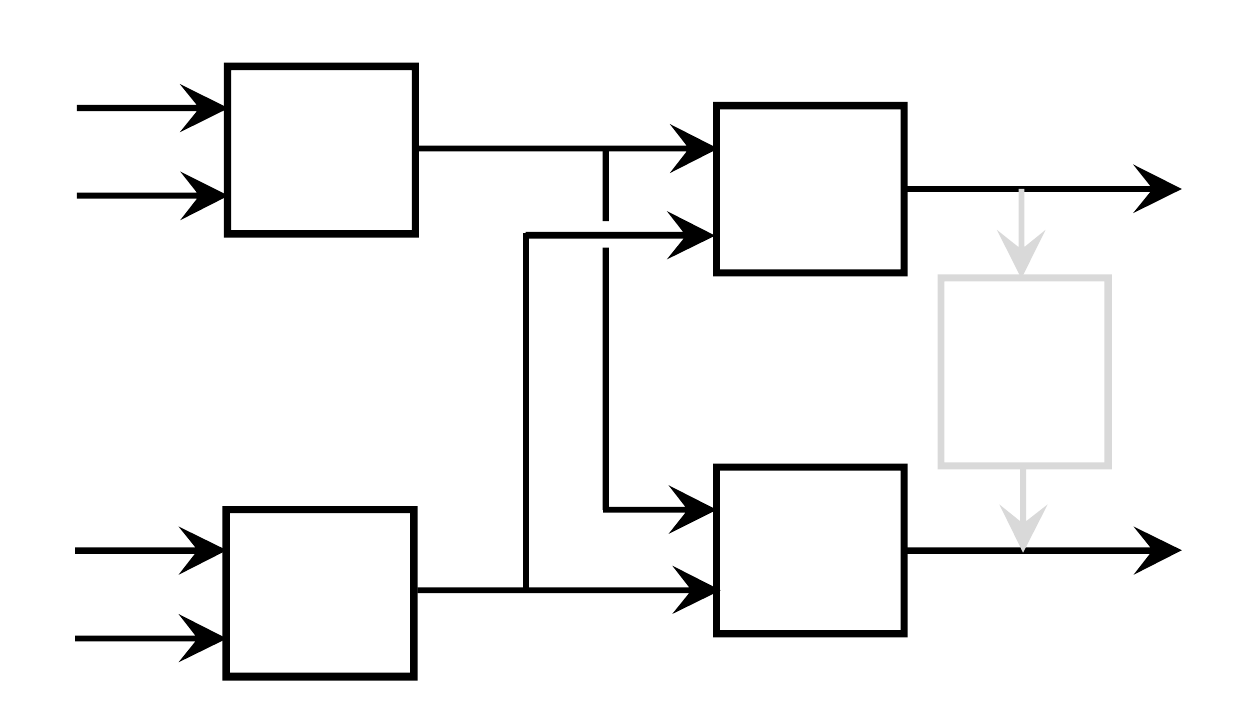}  
 \put (23.3,45)   {$B$}       \put (63, 42)    {$A$}
 \put (23,9)      {$D$}       \put (63,12.5)   {$C$}
 \put(6.6,48) {\color{white}\textbf{///}}
 \put(6.6,40.9) {\color{white}\textbf{///}}
 \put(6.5,12) {\color{white}\textbf{///}}
 \put(6.5,4.9) {\color{white}\textbf{///}}
 \put (81,26)   {\color{gray}$\overline C$}
\end{overpic}
\end{center}
\vspace*{-5mm}
\caption{Node $C$ is $(A,I)$-immanent with model $\overline C$.} 
\label{Fig_3}
\end{figure}

\begin{figure}
\vspace*{-2.5mm}
\begin{center}
\begin{overpic}[width=0.47\textwidth]{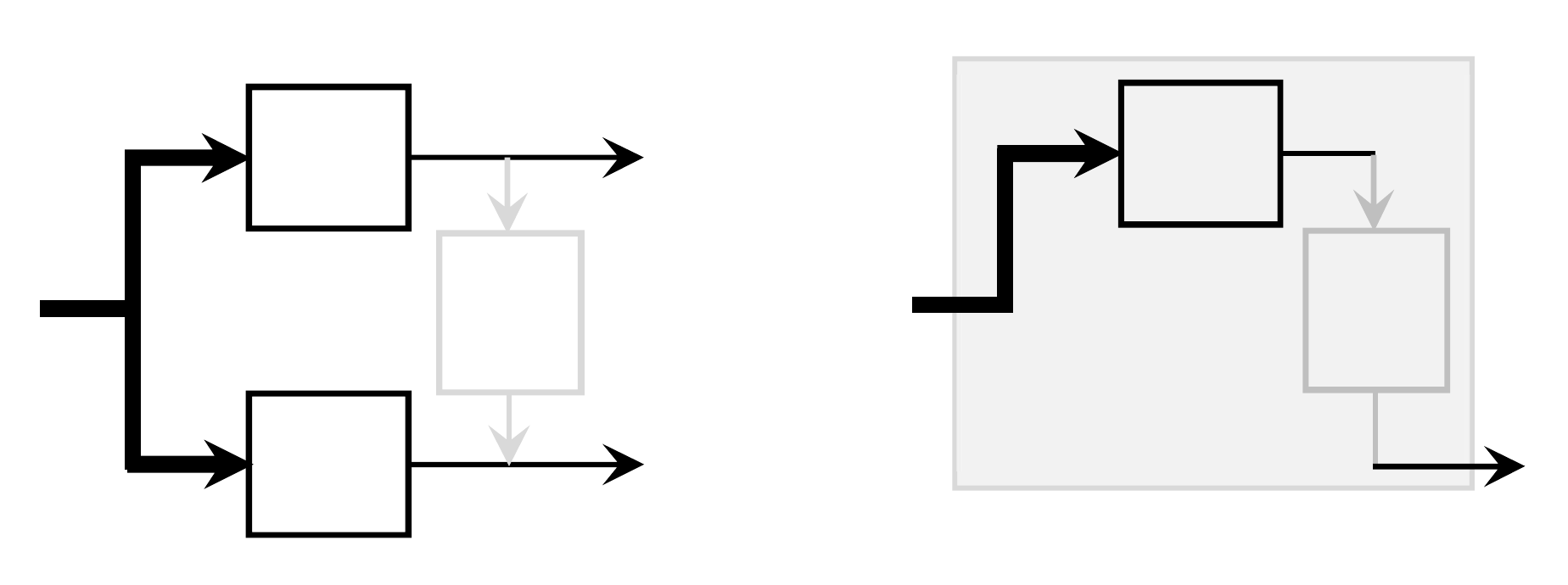}  
 \put (19.2,25.8) {$A$}         \put (74.8, 25.8)  {$A$}
 \put (19.3,6.1)  {$C$}         \put (63,8.5)       {$C$}
 \put (30.9,15.5)   {\color{gray}$\overline C$}
 \put (86.1,15.5) {\color{gray}$\overline C$}

 \put (22,-5.5)     {\scalebox{0.8}{(a)}}
 \put (74,-5.5)     {\scalebox{0.8}{(b)}}
\end{overpic}
\end{center}
\vspace*{-0mm}
\caption{Node $C$ is $(A,I)$-immanent with model $\overline C$.\ \
 (a) Nodes $A$ and $C$ with the common vector input. (b) Node 
 $C=\overline{C}\circ A$ by virtue of Corollary~\ref{ModelingError}.}
\label{Fig_4}
\end{figure}

\begin{figure}
\vspace*{-3mm}
\begin{center}
\begin{overpic}[width=0.48\textwidth]{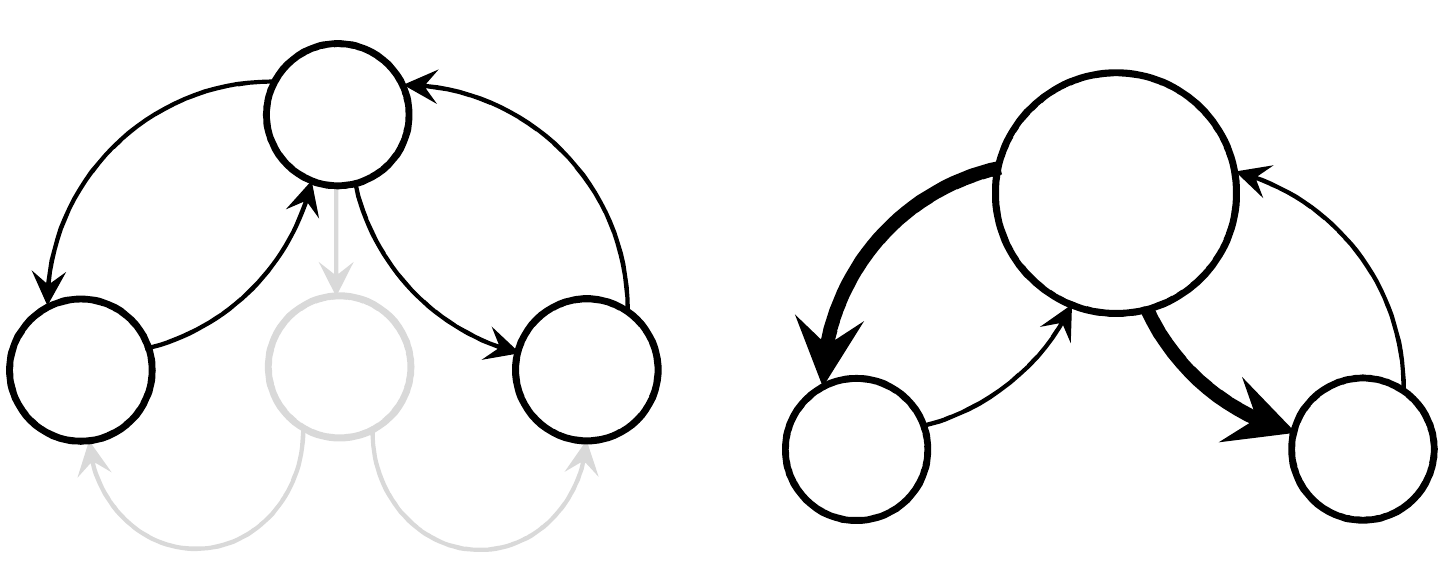}  
 \put (21.4, 31.3)  {$A$}
 \put (38.8,13.4)   {$B$}
 \put (22,13.4)     {\color{gray}$\overline C$}
 \put (3.7,13.4)     {$D$}
 
 \put (57.5,7.9)      {$D$}
 \put (92.6,7.9)   {$B$}
 \put (69.2,26)    
     {\scalebox{.95}{$(_{\!}I\! \oplus\!\overline C)\!\circ_{\!}\! A$}}

 \put (22,-3.3)     {\scalebox{0.8}{(a)}}
 \put (76,-3.3)     {\scalebox{0.8}{(b)}}
\end{overpic}
\end{center}
\vspace*{-0.5mm}
\caption{Reduced network:\ \ (a) With node $C$ represented by
$\overline{C}\circ A$ thus sharing the existing node $A$. (b) With 
the outputs of the new node consolidated.}
\label{Fig_5}
\end{figure}

Unless additional information is available, no  
structural reduction of this network is possible. So, 
suppose that $C$ is $(A,I)$-immanent.  Then, according to 
Corollary~\ref{ModelingError}, there exists a modelling 
mapping $\overline C$, as indicated in Fig.~\ref{Fig_3}.  
Following Lemma~\ref{Incorp}, we can combine the inputs 
feeding into nodes $A$ and $C$ to obtain a common vector 
input feeding into both $A$ and $C$, as depicted by the bold 
line in Fig.~\ref{Fig_4}(a).  Fig.~\ref{Fig_4}(b) shows that 
the mapping $C$ can be replaced by the composition 
$\overline{C}\circ A$ as follows to Corollary~\ref{ModelingError}.
This means that we can now replace node $C$ of the network 
with a ``node'' $\overline C$ which has a single 
input, fed by the output of node $A$.  The output of node 
$\overline C$ then replaces the output of node $C$, 
feeding into nodes $B$ and $D$ as shown in Fig.~\ref{Fig_5}(a).
To reduce this network to a three-node network requires 
us to merge $A$ and $\overline{C}$ into a single node with 
mathematical description $(I\oplus\overline{C})\circ A$ 
resulting in it having a vector output which feeds into 
nodes $B$ and $D$ via the bold edges in the graph shown in 
Fig.~\ref{Fig_5}(b).

Now, if there were no further immanence present in the network, 
then Fig.~\ref{Fig_5}(b) shows the simplest network to which 
the original network can be structurally reduced, using nodal 
rationalization.

\begin{figure}
\vspace*{-4.5mm}
\begin{center}
\begin{overpic}[width=0.49\textwidth]{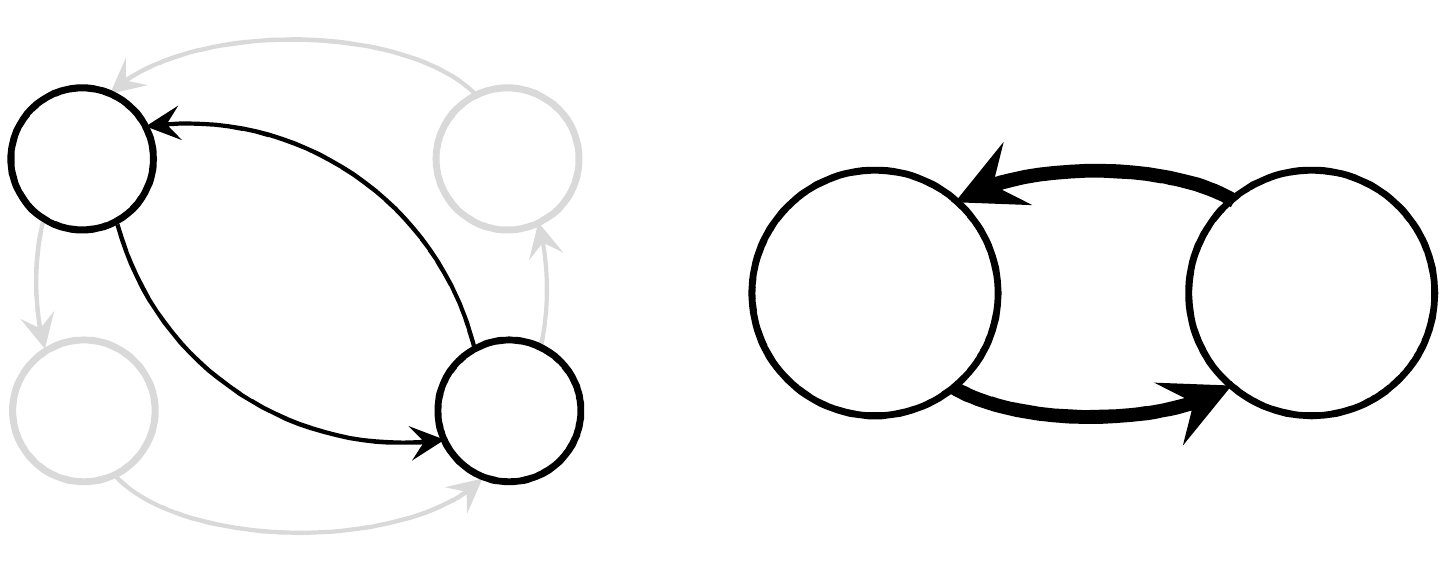}  
 \put (4, 26.5)  {$A$}
 \put (33.6,8.9)   {$B$}
 \put (33.6,26.5)   {\color{gray}$\overline D$}
 \put (4,8.9)   {\color{gray}$\overline C$}

 \put (52.4,17.1)    
     {\scalebox{.95}{$(_{\!}I\! \oplus\!\overline C)\!\circ_{\!}\! A$}}
 \put (82.4,17.0)    
     {\scalebox{.95}{$(_{\!}I\! \oplus\!\overline D)\!\circ_{\!}\! B$}}

 \put (22,-3.3)     {\scalebox{0.8}{(a)}}
 \put (76,-3.3)     {\scalebox{0.8}{(b)}}
\end{overpic}
\end{center}
\vspace*{-1.5mm}
\caption{Reduced network:\ \ (a) With nodes $C$ and $D$ represented by
$\overline{C}\circ A$ and $\overline{D}\circ B$, respectively. (b) With 
the outputs of the new nodes consolidated.}
\label{Fig_6}
\end{figure}
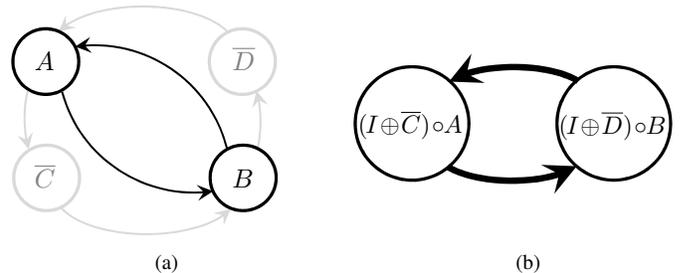

Next, additionally, assume that $D$ is $(B,I)$-immanent.  Once 
again, according to Corollary~\ref{ModelingError}, there exists 
a modelling mapping $\overline D$ such that $D = \overline{D} 
\circ B$.  Following the same procedure as above, the 
network can now be reduced to the form shown in 
Fig.~\ref{Fig_6}(a)---effectively a two-node network as 
shown in Fig.~\ref{Fig_6}(b).  To see 
this simply define the two nodes to have the mathematical 
descriptions\footnote{For economy of presentation and for 
readability, we represent the two identity mappings $I_A$ 
and $I_B$, operating on the ranges of $A$ and $B$, 
respectively, using the same symbol, namely $I$.}
$(I\oplus\overline{C})\circ A$ and $(I\oplus\overline{D})\circ B$, 
respectively, resulting in the interconnecting edges to become 
vector valued. This reduction is striking, considering the 
generality of the mathematical descriptions of the four nodes 
of the original network. 

With the insight developed as our discussion unfolded, we look 
back and point out an important and surprising observation: if our 
example complex network was completely void of immanence, i.e.\ 
all counter-cascaded node pairs were transcendent, then by Theorem~\ref{ModOpTheor}, the original network 
would not in any way have been structurally reducible using 
nodal rationalization.  In other 
words, except perhaps for cosmetic changes, Fig.~\ref{Fig_2} 
would then represent the \emph{simplest} form possible for 
this network and imposing nodal rationalization in such a case, 
would  yield unavoidable and unresolvable approximation 
or modeling errors.

A note about bi-immanence and bi-transcendence is in order: 
for our example here, if bi-immanence was present, then we 
would have had the option to interchange the roles of relevant 
systems, thus giving us more options while still producing the 
same results.

In conclusion we point out that, in this example application, 
we stretched the presence of immanence to the limit in order 
to demonstrate the compactness of representation produced 
by the nodal rationalization.  However, in real life complex 
network investigations, nodal rationalization will usually 
only be applied selectively to expose important latent 
properties that would otherwise have gone unnoticed.

\bigbreak
\section{Conclusion}\label{sec:Conclu}
We presented a theoretical result which states a necessary and 
sufficient condition for multivaluedness when attempting to 
relate two outputs (effects) resulting from the same local or 
global input (cause) in a complex network  Subsequent 
corollaries provide further useful results for determining 
multivaluedness, given special conditions.  

A simple application, namely nodal rationalization, was 
demonstrated, using a simple yet very general four-node 
complex network.  It shows the presented results' potential 
to contribute toward the arsenal of tools for studying 
complex networks and systems in general.

Further work is in progress to specialize these theoretical 
results to applications in distributed measurement in networks, 
big data and neural networks, and in the analysis of signal 
processing algorithms.

\bigbreak
\section*{Acknowledgment}
M.A. van Wyk's research chair in System and Control Engineering is financially supported by the Carl and Emily Fuchs Foundation.

The author wishes to thank both Messrs. A.M.\ McDonald and A.M.\ 
van Wyk for proofreading early drafts of the manuscript and 
Ms. Gina Ng for technical assistance.  

\bigbreak

\end{document}